\documentclass[showpacs,preprintnumbers,amsmath,amssymb,aps,twocolumn,superscriptaddress]{revtex4}

\usepackage{amsfonts}
\usepackage[dvips]{graphicx}
\usepackage{type1cm}
\usepackage{color}
\usepackage{bm}
\usepackage{epstopdf}

\renewcommand{\vec}[1]{\mathbf{#1}}

\newcommand{\nn}{\nonumber \\}

\def\be{\begin{equation}}
\def\ee{\end{equation}}
\def\bea{\begin{eqnarray}}
\def\eea{\end{eqnarray}}

\begin{document}
\title{Hawking Radiation and Non-equilibrium Quantum Critical Current Noise}
\author{Julian Sonner}
\email{js499@cam.ac.uk}
\affiliation{DAMTP, University of Cambridge, CB3 0WA, U.K. \& Imperial College London, SW7 2ZA, U.K. \& KITP, University of California Santa Barbara, CA 93106, U.S.A.}
\author{A.G.~Green}
\affiliation{London Centre for Nanotechnology, University College London, 17-19 Gordon St, London, WC1H 0AH, UK}

\preprint{NSF-KITP-11-243, Imperial/TP/2012/JS/01}


\begin{abstract} 
The dynamical scaling of quantum critical systems in thermal equilibrium may be inherited in the driven steady-state, leading to universal out-of-equilibrium behaviour. This attractive notion has been demonstrated in just a few cases. We demonstrate how holography  - a mapping between the quantum critical system and a gravity dual - provides an illuminating perspective and new results. Non-trivial out-of-equilibrium universality is particularly apparent in current noise, which is dual to Hawking radiation in the gravitational system. We calculate this in a 2-dimensional system driven by a strong in-plane electric field and deduce a universal scaling function interpolating between previously established equilibrium and far-from-equilibrium current noise. Since this applies at all fields, out-of-equilibrium experiments no longer require very high fields for comparison with theory. 
\end{abstract}


\maketitle


Relatively few general principles are known that govern the behaviour of quantum systems driven out of equilibrium \cite{Evans:2012p10720,Evans:2012p10666,PhysRevLett.78.2690,PhysRevE.60.2721}. Those  discovered do not constrain the system as tightly as those of equilibrium thermodynamics. For quantum critical systems, there may be additional principles leading to universal out-of-equilibrium behaviour. Quantum critical systems display dynamical scaling in thermal equilibrium \cite{Sachdev1999}. Since non-equilibrium steady states are largely determined by dynamics, one might anticipate that this  scaling is inherited by the driven steady state. 
This has been seen in a few cases \cite{PhysRevLett.93.027004,PhysRevLett.95.267001,PhysRevLett.97.227003,PhysRevLett.97.236808,PhysRevB.78.195104,Karch:2010kt}. However, analysis of out-of-equilibrium behaviour is fraught with conceptual and calculational difficulties. New insights and approaches are required.

Holography provides one such approach \cite{Herzog:2007ij,springerlink:10.1007/978-3-642-04864-7_9}. It utilizes a mapping between gauge theory in $d$ dimensions and string theory in $d+1$ dimensions; the classical gravity saddlepoint of the string theory corresponding to a strongly-correlated quantum state of the gauge theory. 
In this way, holography encodes quantum many-body correlations in a classical, gravitational metric, allowing complementary  physical insights and calculational tools to be brought to bear \cite{Herzog:2007ij,springerlink:10.1007/978-3-642-04864-7_9,PhysRevLett.105.151602,Faulkner:2010zz,Cubrovic:2009ye,Hartnoll:2009ns,Hartnoll:2007ih}. 

Here, we apply the holographic technique to the study of non-linear current noise - dual to Hawking radiation - at a two-dimensional $z=1$ quantum critical point driven by an in-plane electric field. 
 We find a noise power 
 \begin{equation}
 S_j \sim 4 \sigma T_*
 \;\;\; \hbox{with} \;\;\;\;
\pi k_B T_* = [ (\pi k_B T)^4+\hbar^4 c^4 e^2E^2]^{1/4}
\label{MainResult}
\end{equation} 
interpolating between Johnson-Nyquist noise at low fields and previous results at high-fields \cite{PhysRevLett.97.227003}.
Whilst driven out-of-equilibrium steady-state distributions are generally very different from those in equilibrium, holography reveals that the out-of-equilibrium {\it fluctuations} of a quantum critical system may be {\it precisely thermal}.

 \begin{figure}[h!]
\begin{center}
 \includegraphics[width=2.2in]{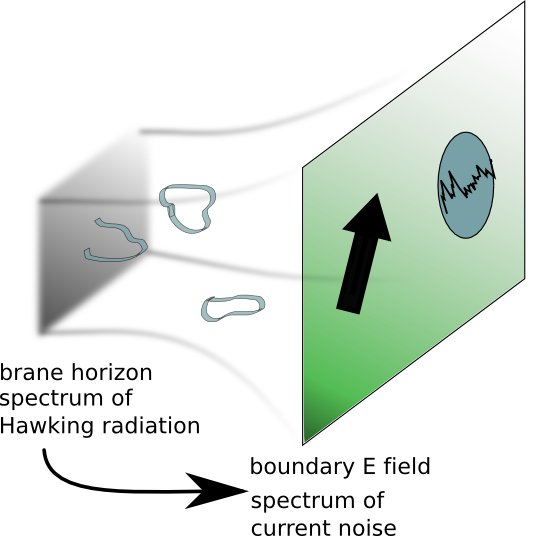} 
\caption{\it Sketch of the AdS geometry with boundary current fluctuations: The AdS brane geometry is modified by the application of an in-plane electric field at the boundary. Hawking radiation from the modified horizon propagates to the boundary where it induces current fluctuations. These fluctuations satisfy an equilibrium-like fluctuation dissipation relation at a temperature $T_* = [T^4+\hbar^4 c^4 e^2E^2/(\pi k_B)^4]^{1/4}$, interpolating between established low- and high-field limits.}
\end{center}
\end{figure}
Our calculation is carried out in a probe-brane limit with boundary conditions corresponding to the application of the in-plane electric field \cite{Karch:2010kt}. This field induces a modified 
metric \cite{Kim:2011qh} on the probe brane with a horizon at temperature $T_*$ \footnote{Thermal spectral functions exhibiting an effective temperature were previously noticed  in the context of quark diffusion, see for example \cite{CasalderreySolana:2007qw,Gubser:2006nz,Gursoy:2010aa}.}. 
We show explicitly that Hawking radiation from the emergent horizon generates current fluctuations in the boundary that are thermal at temperature  $T_*$ (see Fig. 1), a conclusion consistent with general gravitational principles and with potentially profound implications for out-of-equilibrium quantum criticality.

These results are compared to those for the bosonic superconductor to insulator transition, embodied in the Bose-Hubbard model. This is amongst the quantum critical systems that we have under the best analytical control \cite{PhysRevB.44.6883,PhysRevB.56.8714}, making it ideal  to test new ideas and techniques. Early work on out-of-equilibrium 
quantum criticality considered this system \cite{PhysRevLett.95.267001,PhysRevLett.97.227003}. Pioneering holographic studies have reproduced dualities of the equilibrium magneto-thermo-electric response\cite{Hartnoll:2007ih,PhysRevLett.98.166801} and the steady-state out-of-equilibrium current \cite{PhysRevLett.95.267001,Karch:2010kt}. These results are a useful benchmark and demonstrate the utility of the holographic approach. 

We begin by outlining how the non-equilibrium steady-state response and current noise are obtained in a $1/N$-expansion of the Bose-Hubbard model. We follow with a heuristic discussion of the holographic mapping and how it can be applied to out-of-equilibrium steady-states using the probe brane limit. We pay particular attention to the similarities between the trick used in the $1/N$ expansion and the probe brane limit. Next, we detail the calculation on the gravity side and end with a critical discussion of the results and their implications.

{\it Out-of-Equilibrium Noise in the Bose-Hubbard Model:}
The continuum field theory describing the two-dimensional Bose-Hubbard model is essentially a Klein-Gordon model with a $\phi^4$ interaction, minimally coupled to an in-plane electric field \cite{PhysRevB.44.6883}. The Klein-Gordon field supports normal modes of positive and negative charge. 
Charge transport in this model can be described using a Boltzmann equation for the population of positively and negatively charged normal modes \cite{PhysRevB.56.8714}, controlling the interaction within an $\epsilon$-expansion. The presence of both charges allows  scattering to relax current even though it conserves both energy and momentum. Formally, the model has infinite thermal conductivity, $\kappa$, at leading order in $\epsilon$ and finite electrical conductivity, $\sigma$.

As the electric field is increased and the system driven away from linear response, new processes become important and account must be taken of heat flow. The strong electric field produces pairs of positive and negative modes from the vacuum by the Schwinger mechanism/Landau-Zener tunnelling \cite{PhysRevLett.95.267001}. At the same time, Joule heating pumps energy into the system. In order to achieve a steady state, there must be a heat sink.
One way is to consider a finite system so that excess heat and particles can be transported to the boundaries. As emphasized by Tremblay \cite{PhysRevA.19.1721}, heat transport requires gradients in free-energy density. If this variation  is to be small, and the response independent of the size and geometry of the sample, there is a limit on size. However, for particle-hole symmetric systems, the Wiedemann-Franz law does not hold and thermal conductivity can be much larger (in appropriate units) than electrical conductivity. In this case, a very small thermal gradient is required to support the flow of heat to the edge of the sample and maintain a steady state. Indeed, for the Bose-Hubbard model, the leading thermal conductivity is infinite and the system size may be taken to infinity provided that we take $\kappa$  to infinity first \cite{PhysRevLett.98.166801}.
Both the $1/N$ expansion and the probe brane calculation take advantage of these arguments for the existence of a steady-state out-of-equilibrium distribution and calculate the  out-of-equilibrium distribution directly for an infinite system.

 In the $1/N$-expansion \cite{PhysRevLett.97.227003}, the electric field is coupled to just one of the components of an $O(N)$ vector field. The $N-1$ components that are not coupled to the electric field effectively act as a critical bath for the field-coupled component. This 
 allows the steady-state out-of-equilibrium conductivity to be calculated in a spatially uniform, infinite system.
A Boltzmann equation is developed for the field-coupled component of the $O(N)$ field. 
Integrating over momentum, one recovers a Boltzmann equation for the total current;
$$
\frac{d j}{dt} = aE^{\frac{d+1}{2}}-b \sqrt{E} j,
$$
where the first term on the right hand side has its origin in Schwinger pair production and the second term describes the decay of current carried by the field-coupled component to the remaining $N-1$ components. The constants $a$ and $b$ characterise these two processes\footnote{We adopt the same notation here as in Karch and Sondhi.}. Integrating this equation, one obtains a steady-state current 
$j=\frac{a}{b} E^{d/2}$ 
giving a conductivity 
$\sigma = \frac{a}{b} $
in two dimensions. Note that in this $1/N$ treatment, the steady-state distribution is {\it not} simply thermal, a point that we shall return to later.

The current noise is determined  by developing a Boltzmann-Langevin \cite{JETP1969,PhysRev.187.267} description of fluctuations in the distribution function. The underlying physical insight is that the scattering processes may be treated as independent and so obey  Poisson statistics. 
Fluctuations in the distribution function may then be approximated by adding Gaussian noise to the Boltzmann equation with variance proportional to the mean scattering rate. 
Integrating this Boltzmann-Langevin equation results in the following equation for the current fluctuations:
\begin{eqnarray*}
\frac{d \delta j}{dt}& =& -b \sqrt{E} \delta j +\eta,
\\
\langle \eta(t) \eta(t') \rangle &=& c E^{(d+1)/2} \delta(t-t').
\end{eqnarray*}
Integrating this equation and considering the behaviour on long timescales, one obtains the large-$E$ (zero temperature) limit of the non-equilibrium current noise given by Eq.(\ref{MainResult}).

{\it Heuristic Picture of the AdS Realization:}
As discussed above, holography uses a mapping between a critical gauge theory in $d$ dimensions and a string theory in $d+1$. The gauge theory may be thought of as describing or living on the boundary of the gravitational system with the extra dimension playing the role of a renormalization group scale.
The classical gravity saddlepoint of the string theory encodes a strongly-correlated quantum state of the gauge theory. In particular, the interactions and scaling are embodied in the metric. The horizon in the presence of a black hole has particular importance in what follows. As usual, its area determines the entropy of the black hole, whose temperature is given by the surface gravity. This temperature has the same interpretation in both the gauge theory and  gravity.

We model non-linear transport using the $D3_N/D5_M$ brane intersection \cite{DeWolfe:2001pq,Erdmenger:2002ex}: we study the dynamics of $M$ $D5$ branes in the background of $N D3$ branes in the probe brane limit $N \gg M$. The specific configuration is summarized in table \ref{tab:BraneSetup}. The D3 branes lead to the formation of an horizon in the background metric modelling the effect of a thermal bath. Non-linear transport is studied through the dynamics of gauge fields on the D5 brane with boundary conditions corresponding to the application of an in-plane electric field.
\begin{table}[htdp]
\caption{{\bf The $D3_N/D5_M$ brane intersection}. We label directions occupied by the branes by $\bullet$ and directions in which they are localised by $\circ$. The field-theory directions are $x^0,x^1,x^2 \equiv t,x,y$. In addition the $D5$ wraps the RG direction $x^4\equiv u$ as well as an $S^2 \subset S^5$, parametrised by two angles $x^5,x^6\equiv \theta_1,\phi_1$. This system realises an out-of-equilibrium steady state in $2+1$ dimensions. 
\label{tab:BraneSetup}}
\begin{center}
\begin{tabular}{c||c|c|c|c|c|c|c|c|c|c||}
& 0 & 1 & 2 & 3 & 4 & 5 & 6 & 7 & 8 & 9 \\
\hline\hline
D3 & $\bullet$ & $\bullet$& $\bullet$ & $\bullet$ & $\circ$& $\circ$ &$\circ$ &$\circ$&$\circ$&$\circ$\\
D5 & $\bullet$ & $\bullet$ & $\bullet$ &$\circ$ & $\bullet$ &$\bullet$ &$\bullet$ &$\circ$ &$\circ$ & $\circ$
\end{tabular}
\end{center}
\label{default}
\end{table}
This permits us to study out-of-equilibrium steady states in a very similar way to the $1/N$ trick used in the condensed matter calculations of current noise. Since $N \gg M$, gauge fields on the D5 brane do not modify the metric. Just as the single component of the $O(N)$ field coupled to the electric field does not modify the distribution of the $N-1$ modes corresponding to the critical bath from the condensed matter point of view.

The application of the electric field on the boundary leads to a crucial modification of the metric induced on the D5 brane. A new horizon appears in this reduced metric with an area that  corresponds to the effective temperature $T_{*}$ in Eq.(\ref{MainResult}). A major result of this is that the conductivity and current fluctuations are related 
in a way that mimics the equilibrium fluctuation-dissipation relation at the effective temperature $T_{*}$. Such a relation 
appears to be inevitable in the gravity dual. As we discuss below, the consequences of this are potentially profound.

{\it Details of the AdS Calculation:} Our calculation, involves solving the equations of motion for gauge fields on the D5 brane. The steady-state solution corresponds to a previous calculation that allows determination of the out-of-equilibrium conductivity \cite{Karch:2010kt}. We explicitly propagate thermal fluctuations from the horizon on the D5 brane to its boundary and derive a Langevin equation for gauge fluctuations and hence current fluctuations on the boundary. 
The dynamics of the $D3_N/D5_M$ brane intersection is captured by the action
\be\label{eq:D5Action}
S_{D_5} = -{\cal N}_5\int d^6\xi \sqrt{-det\left(g_{\rm ind} + F  \right)} + S_{\rm WZ}\,,
\ee
where $g_{\rm ind}$ is the metric induced on the $D5$ brane, and $F$ is the gauge field strength on the brane in string units. We do not specify the Wess-Zumino term $S_{\rm WZ}$ further as it does not contribute to the configurations that we study below. The normalisation ${\cal N}_5 = M T_5$ where $T_5$ is the usual $D5$-brane tension.
We study this system in the background of the non-extremal $D3$ brane metric, which we take to be
\be
G_{mn}dx^mdx^n= \frac{u^2}{R^2} \left(-f(u) dt^2 + d\vec{x}^2  \right) + \frac{R^2du^2}{f(u) u^2} + R^2 d\Omega_5^2
\ee
with $f(u) = 1 - \frac{u_h^4}{u^4}\,,$
and $d\Omega_5^2$ is the metric on a round five sphere. By studying the regularity of the Euclidean geometry at $u_h$ or otherwise, one sees that the background geometry (the {\it bath}) has a temperature
$T = u_h/(\pi R^2)$.

{\it Applying an external electric field} in the $x$ direction is accomplished by choosing the gauge potential
\be
A = \left( -\tilde E t - A_x(u)  \right) dx\,,
\ee
where $\tilde E$ is a dimensionless electric field.
The geometry of the brane is obtained by solving the  equations resulting from the action (\ref{eq:D5Action}). We choose the static gauge on the brane. 
The degrees of freedom are a single angle $\theta(u)$ describing the position of the $S^2$ on the $S^5$ and the radially dependent function $A_x(u)$. An important simplification allows us to set $\theta(u)=0$, corresponding to the case of coupling {\it massless} charged degrees of freedom to the field. 

The field $A_x(u)$ is a cyclic coordinate and hence has  first integral
\be\label{eq:MatchingCondition}
A_x'(u) = C \sqrt{\frac{G_{uu} \left(  \tilde E^2 + G_{tt}G_{xx}  \right)}{G_{tt} \left( C^2 + G_{yy} G_{tt}  \right)}}\,,
\ee
where $C$ is an integration constant related to the current $C = j/{\cal N}_5$. Requiring that the expression be real for all values of $u$ demands \cite{Karch:2007pd} that denominator and numerator cross zero at the same point $u:=u_*$. This determines $u_*^4 = u_h^4 + \tilde E^2 R^4$. The special point $u_*$ on the brane coincides with the horizon of the background black hole iff $\tilde E=0$. This matching procedure also relates the current to the applied field, yielding (see \cite{Karch:2010kt})
\be\label{eq:Conductivity}
j = {\cal N}_5 \tilde E =: \sigma_{(2+1)}\,\tilde E\,,
\ee
{\it i.e.} non-linear conductivity in $d=3$ is a constant \cite{PhysRevLett.95.267001,PhysRevLett.93.027004}.

{\it Brane Horizon, Effective Temperature:}
This result, and its higher-dimensional generalisations, can further be interpreted in terms of an effective geometry on the brane, seen by the fluctuations of the gauge field.  Let us perturb $A\rightarrow A +  a$, where
\be
a = a^\|(u,t) dx + a^\perp(u,t) dy + a^t(u,t) dt\,.
\ee
Then the quadratic fluctuation action becomes
\begin{widetext}
\bea
S^{(2)} &=& \tilde{{\cal N}_5}\int du dt
 \sqrt{-\alpha} \alpha^{ab}\left(\partial_a a^\| \partial_b a^\|  + Z(u)^\perp\partial_a a^\| \partial_b a^\|  + Z^t(u) \partial_a a^t \partial_b a^t \right)\,
\eea
\end{widetext}
where $\tilde{\cal N}_5$ results from integrating out the angular directions, and for the case of no spatial dependence also the $x,y$ directions. $Z^I(u)$ are conformal rescaling factors whose exact expressions we do not need. Upon plugging in the conductivity (\ref{eq:Conductivity}) we find that $Z^\perp(u)=1$.
 By choosing a new time variable $\tau$ we can put the $\tau,u$ part into diagonal form
\be\label{eq:OSM}
ds^2 = -\frac{u^4-u_*^4}{u^2 R^2}d\tau^2 + \frac{u^2R^2du^2}{u^4 - u_*^4}\,,
\ee
where $\tau = t + f(u)$ with
\bea
f(u) &=& \frac{1}{2} 
\left[ 
\frac{1}{u_h}\left(\arctan\frac{u}{u_h}  -  {\rm arctanh}\frac{u}{u_h} \right)  
\right.
\nonumber\\
& & \;\;\;\;\;\; \left.+ \frac{1}{u_*}\left(\arctan\frac{u}{u_*}  -  {\rm arctanh}\frac{u}{u_*} \right)  \right] \,,
\eea
which 
is akin to the trailing string solution. The fluctuations for $u>u_*$ are causally disconnected from the region $u<u_*$, so that $u_*$ has precisely the form of a black-hole horizon. Euclidean regularity of this horizon imposes the temperature
\bea
\pi T_* &=& \frac{u_*}{R^2}  = \left[ \tilde E^2 R^{-4}+ \left( \pi T \right)^4  \right]^{1/4}.
\eea
Restoring units \footnote{This is done in the standard way via $\tilde E = \alpha'  E$ so that $\tilde E^2 R^{-4} = \tfrac{\alpha'^2}{R^4} E^2 \sim 1/\lambda E$, where $ \alpha'$ is the string length and $\lambda$ is the 't Hooft coupling of the dual field theory. } and assigning charge $e=1/\sqrt{\lambda}$ to the fundamental charge carriers, as in \cite{Karch:2007pd,Karch:2010kt}, this may be expressed in terms of universal field theory quantities as in Eq.(\ref{MainResult}).
This effective temperature  interpolates precisely between the two limits $T_*\sim\sqrt{E}$ and $T_*\sim T$ which emerged from earlier calculations near the superconductor-insulator transition.  Furthermore, the spatial fluctuations $\mathbf{a} = (a^\|,a^\perp)$ decouple from $a^t$ so that we are free to consider them separately. The fluctuations of interest satisfy the bulk equation
\be\label{eq:SpatialFluctuations}
\frac{\frak{w}^2}{z^2 f(z)}\mathbf{a}_i + \partial_z \left( z^2 f(z)\partial_z \mathbf{a}_i \right)=0\,,
\ee
where we have changed to the dimensionless variable $z = u/u_*$ and $f(z) = 1-z^{-4}$. The quantity $\frak{w}:=\omega/(\pi T_*)$ is a dimensionless measure of frequency.
We can now study the noise of the current $\mathbf{j}$, dual to the bulk gauge field $\mathbf{a}$, by following the prescription of \cite{Herzog:2002pc} to calculate the Schwinger-Keldysh two-time correlators of the boundary field theory. The current fluctuations satisfy a Langevin equation of the form
\be
\frac{d\mathbf{j}}{dt} + \int  \gamma(t,t')\mathbf{j}(t')dt' = \xi(t)\,,
\ee
where we can extract the memory kernel $\gamma(t,t') = G_R(t,t')$ and noise $\xi(t)$ using holography, in analogy to holographic Brownian motion of heavy quarks \cite{deBoer:2008gu,Son:2009vu}.

 By carefully studying the analyticity properties of positive frequency modes on the Kruskal plane as we cross the horizon \cite{Herzog:2002pc}, we find the noise correlation
\be\label{eq:FluctDiss}
\langle \xi^i(\omega)\xi^j(-\omega)\rangle = G_{\rm sym}^{ij}(\omega) = -(1 + 2 n_*){\rm Im} G_R^{ij}(\omega)\,,
\ee
where $G_{\rm sym}^{ij}(\omega)$ is the symmetric/Keldysh correlator of the current $\mathbf{j}_i$ in the Schwinger-Keldysh formalism and $G_R^{ij}(\omega)$ is the retarded correlator. We have the effective thermal factor $n_* = 1/(\exp (\omega/T_*)-1)$. This gives us a current noise
\bea
S_j &=& - \int_{-\infty}^\infty d\omega dte^{i\omega t}{\rm coth}(\omega/2T_*) {\rm Im}\langle \mathbf{j}(\omega) \cdot \mathbf{j}(-\omega)  \rangle_R \nn
&=& -  T_* \lim_{\omega\rightarrow 0} \frac{2}{\omega}{\rm Im}\,{\rm Tr}G^R(\omega) = T_* \Phi(T/T_*) \,,
\eea
where the final equality follows from a scaling analysis of
 Eq. (\ref{eq:SpatialFluctuations}). In fact, we can solve equation (\ref{eq:SpatialFluctuations}) exactly. The solution with ingoing boundary conditions at the horizon for both components of $\mathbf{a}$ is 
\be
\mathbf{a}_i(z) = c_i\left(\frac{z-1}{z+1}\right)^{-\frac{i\frak{w}}{4}}\exp\left({-\frac{i\frak{w}}{2}\arctan z}\right)\,,
\ee
for constants $c_i$. From this we deduce the exact retarded Green function
\be
G_R^{ij}(\omega) = -i \sigma_{(2+1)}\frak{w}\delta^{ij}\,,
\ee
where $\sigma_{(2+1)}$ is the non-equilibrium conductivity determined
above.
In turn this gives the current noise power
\be
S_j = 4 \sigma_{(2+1)}T_*\,.
\ee
The noise is related to the conductivity by an  equilibrium-like  fluctuation-dissipation or Einstein relation at the effective temperature; an exact interpolation between the previously known results of \cite{PhysRevLett.97.227003}.

{\it Discussion:}
Hawking radiation is dual to quantum critical current noise under the holographic mapping. We have used this mapping in the probe brane limit to study non-linear current noise. The crucial feature of this analysis is the effect of the electric field on the metric induced on the probe brane; this metric is modified such that a new horizon appears with which we may associate a temperature 
 $T_*$\footnote{Note that one cannot associate an equilibrium temperature to the overall Euclidean metric, as there will be a conical deficit either at $u_h$ or at $u_*$. This is a clear sign that the system as a whole is far from equilibrium.}. 
This horizon describes the combined effects of thermal fluctuations from the equilibrium bath and pair production by the Schwinger mechanism. 
By considering the propagation of fluctuations from this horizon to the boundary, we have shown that the fluctuations on the boundary - and hence of the gauge theory - obey an  equilibrium-like fluctuation-dissipation or Einstein relation at the effective temperature $T_*$. 

In general, an out-of-equilibrium steady state is not expected to obey an equilibrium-like fluctuation dissipation relation. The present holographic analysis, suggests that this may nevertheless be the case in quantum critical systems with a holographic dual. This rather elegant result emerges as a consequence of some very general properties of gravitational metrics. It would be very informative to re-cast it as a principle directly applicable to the condensed matter system. This principle is subtly hidden; 
it does not correspond to a thermal distribution of the Klein-Gordon normal modes, since this would not carry a current. The distribution of fluctuations is a combined feature of the non-equilibrium distribution and the scattering integral. 
Nevertheless, thermal distributions of fluctuations in the presence of profoundly non-equilibrium distributions have been seen before in quantum critical systems in the low frequency limit \cite{PhysRevLett.103.206401,KirchnerQi}, even when they do not posses a known holographic dual. One might then speculate that if this principle can be identified, it will have broader application.

It would seem that for a quantum critical system with a gravity dual, we may always consider non-linear response in the probe brane limit. Presumably, the driving field will always modify the induced metric in such a way that a new horizon is formed with which we may associate an effective temperature. As we have taken pains to demonstrate here, a fluctuation-dissipation relation at this effective temperature will result.  What are the limitations of this picture? We have used arguments from condensed matter to motivate the existence of a non-equilibrium steady state in the correct limit of thermal conductivity and system size. These conditions are not easily met in practice and certainly constrain the applicability of our results. Nevertheless, where they are met, it is unlikely that a gauge-gravity dual that applies in equilibrium would be invalidated by simply tuning up the strength of the probe field.
One way to negotiate the constraints of system size and thermal conductivity, and an interesting avenue for further study, might be to study the approach to the steady state distribution in a quench after the sudden change of electric field similar to the mass and coupling quenches of \cite{Das:2010yw}.

{\it Acknowledgments:} We thank F. Dowker, C. Herzog, U. Gursoy, S. Hartnoll, A. Karch, J. McGreevy, R. Myers, H. Ooguri, P. Phillips, O. Saremi, D. Tong and J. Zaanen for enjoyable discussions. This research was  supported in part by the National Science Foundation under Grant No. PHY05-51164 and EPSRC  under grant code EP/I004831/1.

\bibliography{PRL}{}
\bibliographystyle{prsty}

\end{document}